# Hybrid Information Retrieval Model For Web Images

Youssef Bassil

*LACSC – Lebanese Association for Computational Sciences*
*Registered under No. 957, 2011, Beirut, Lebanon*
youssef.bassil@lacsc.org

*Abstract*— **The Bing Bang of the Internet in the early 90's increased dramatically the number of images being distributed and shared over the web. As a result, image information retrieval systems were developed to index and retrieve image files spread over the Internet. Most of these systems are keyword-based which search for images based on their textual metadata; and thus, they are imprecise as it is vague to describe an image with a human language. Besides, there exist the content-based image retrieval systems which search for images based on their visual information. However, content-based type systems are still immature and not that effective as they suffer from low retrieval recall/precision rate. This paper proposes a new hybrid image information retrieval model for indexing and retrieving web images published in *HTML* documents. The distinguishing mark of the proposed model is that it is based on both graphical content and textual metadata. The graphical content is denoted by color features and color histogram of the image; while textual metadata are denoted by the terms that surround the image in the *HTML* document, more particularly, the terms that appear in the tags *<p>*, *<h1>*, and *<h2>*, in addition to the terms that appear in the image's *alt* attribute, *filename*, and *class-label*. Moreover, this paper presents a new term weighting scheme called *VTF-IDF* short for Variable Term Frequency-Inverse Document Frequency which unlike traditional schemes, it exploits the *HTML* tag structure and assigns an extra bonus weight for terms that appear within certain particular *HTML* tags that are correlated to the semantics of the image. Experiments conducted to evaluate the proposed *IR* model showed a high retrieval precision rate that outpaced other current models. As future work, the proposed model is to be extended to support not only web images but also web videos and audio clips, as well as other types of multimedia files.**

***Keywords***— *Image Information Retrieval, CBIR, Term Weighting, Web Images, HTML*

## 1. Introduction

The idea of searching for information was first mentioned by Vannevar Bush in 1945 [1]. Ten years later, the first operational information retrieval (*IR*) system was born. Its sole purpose was to index text documents and then to allow computer operators to search for information within these documents. In fact, after a long period of research and development, the applications of *IR* became diverse; they include but not limited to document indexing and archiving, information extraction, knowledge storage, document classification and clustering, question-answering, speech retrieval, and web searching.

The advent of the World Wide Web has led to a tremendously massive distribution of data over the Internet. Moreover, as web browsers became multimedia-enabled, web images became abundant over the web. As a result, several image *IR* systems were conceived to handle the storage and retrieval of web images. However, most of them are keyword-based systems in that they index, store, and retrieve images based on their textual metadata such as the image's filename and several *HTML* attributes, and not based on the graphical characteristics depicted in the image itself [2]. These textual retrieval models have common drawbacks; they are laborious as manual annotation of images requires extra time and effort, and inaccurate as it is very difficult to describe images with human natural languages. On the other hand, *CBIR*, short for Content-Based Image Retrieval is a type of image retrieval model that analyzes and processes the content of an image so as to extract visual information out of it [3]. In particular, visual information can refer to colors, textures, shapes, or any sort of graphical attributes associated with the image.

This paper presents a novel hybrid *IR* model for the indexing and retrieval of web images stored within *HTML* documents. Fundamentally, the proposed model indexes web images based on their content as in *CBIR* systems and on textual metadata as in keyword-based *IR* systems. The image's color features are used to describe the visual content of the image; while the text that surrounds the image is used to describe the textual metadata. During the indexing process, the color histogram of the web image is constructed. Then its Euclidean distance to other color histograms from the Corel image database [4] is computed. The Corel image database contains 1000 images organized into 10 different classes according to their color distribution, useful in solving color-based image classification problems. Consequently, the image being indexed is classified by assigning it a *class-label* identical to the *class-label* of the image in the Corel database whose color histogram has the closest Euclidean distance. Afterwards, textual metadata are extracted from the surrounding text of the image, mainly terms enclosed in tags *<p>*, *<h1>*, and *<h2>*, and terms that are part of the image's *alt* attribute and its *filename*. Additionally, the already deduced image's *class-label* is added to the collection of the extracted metadata.

The last step is to weight the extracted metadata and encode them into a vector space model (*VSM*). For this reason, a new term weighting scheme, namely *VTF-IDF* short for Variable Term Frequency-Inverse Document Frequency is used. Characteristically, the *VTF-IDF* assigns a variable weight for the different metadata terms depending on their original source, that is, based on the tags (either *<p>*, *<h1>*, or *<h2>)* and other attributes (either *alt*, *filename*, or *class-label*) in which they were initially enclosed. This is unlike other traditional weighting schemes which treat all terms unvaryingly and assign them equal weights.

The proposed model is expected to improve the precision-recall rate of image retrieval as it combines both the benefits of keyword-based and content-based *IR* models.

## 2. State-of-the-Art

In an information retrieval context, every document is modeled into what so-called a vector space model (*VSM*) [5] which contains several dimensions each corresponding to a particular term weight that represents the frequency of a term in the document. Inherently, terms or words that have occurred frequently in a text should have a higher weight than other terms [6]. This indicates how much a term is significant with respect to its context [7]. For instance, a web document containing five terms *door*, *child*, *tree*, *bridge*, and *weather* with term frequencies 7, 3, 1, 2, and 8 respectively, is represented using *VSM* as $d = (7,3,1,2,8)$. In general, a feature vector for a document $d_j$ is represented as $d_j = (w_{1j}, w_{2j}, w_{3j}, \cdots, w_{nj})$ where $d_j$ denotes a particular document, $n$ denotes the number of distinct terms in document $d_j$, and $w_{ij}$ denotes the weight for the $i_{th}$ term in document $d_j$.

### 2.1 Traditional Term Weighting Schemes

Different term weighting schemes exist, the best known are the Boolean scheme [8], Term Frequency (*TF*) [9], Inverse Document Frequency (*IDF*) [10], and Term Frequency-Inverse Document Frequency (*TF-IDF*) [6].

#### *2.1.1 Boolean Term Weighing*

It is also called binary weighing because it assigns binary weights for terms based on their presence or absence in the document [8]; a term is assigned a value of 1 if it appears at least one time in the document, and is assigned a value of 0 if it never appears. More generally, binary weights can be computed using the following equation:

$$\chi(t) = \begin{cases} 1 & \text{if } t > 0 \\ 0 & \text{if } t = 0 \end{cases}$$

Apparently, the Boolean method is not suitable for document weighting, it more fits query weighting as it does not distinguish between terms that appear one time and terms that appear frequently.

#### *2.1.2 Term Frequency (TF)*

The raw frequency of a term within a document is called term frequency, and is defined by $tf_{t,d}$ with subscripts $t$ denoting a particular term and $d$ denoting a particular document [9]. Put simply, *TF* is the number of occurrence (frequency) of a particular term in a document. *TF* usually reflects the notion that terms occurring frequently in a document may be of greater importance than terms occurring less frequently, and therefore should have a heavier weight. Although *TF* is still considered to some extent an efficient approach to weight terms due to its simplicity and efficiency, it gives too much credit for terms that appear frequently and hence it may discriminate less important documents over more important documents [11].

#### *2.1.3 Inverse Document Frequency (IDF)*

The purpose of *IDF* is to attenuate the weight for terms that occur too often in a collection of documents and to increase the weight for terms that occur infrequently [10]. In other words, rare terms have high *IDF* and common terms have low *IDF*. For instance, a term such as "the" is likely to have a very low *IDF* somewhat close to 0 because it usually appears in every document, while a term such as "psychophysics" is likely to have a very high *IDF* because it appears in very few documents. *IDF* is generally defined as:

$$idf_j = \log\left(\frac{n}{df_j}\right)$$

where $idf_j$ is the Inverse Document Frequency, $df_j$ is the number of documents in which term *j* occurs, and *n* is the total number of documents in the collection. A *log* function is usually used to reduce the value of $idf_j$ especially when the collection includes a huge number of documents.

### *2.1.4 TF-IDF*

Term Frequency-Inverse Document Frequency (*TF-IDF*) weights a particular term by calculating the product of its term frequency (*TF*) in a document with the *log* of its inverse document frequency (*IDF*) in the collection, and thereby discriminating terms that are frequent in a particular document but globally rare in the collection. In other words, terms that are frequent in a given document and infrequent in the whole collection are assigned high *TF-IDF* weight. *TF-IDF* is generally defined as:

$$w_{ij} = tf_{ij} \cdot idf_j$$

where $w_{ij}$ is the final weight for term *j*, $tf_{ij}$ is the frequency of term *j* in document *i*, and $idf_j$ is the Inverse Document Frequency of term *j*. *TF-IDF* is by far the most successful document term weighting scheme and is applicable to almost all vector space information retrieval systems [6].

## 2.2 CBIR Models

Several systems have been developed to index and retrieve web images based on their content. Major ones include the MIT PhotoBook [12], IBM QBIC [13], and the Virage Image Search Engine [14]. Basically, *CBIR* systems capture and extract visual features out of the image and index them appropriately into relational databases. These visual features fall into three categories: color, texture, and shape features.

### *2.2.1 Color Features*

Color features are the easiest to obtain and the most widely used visual attributes. They are usually the color intensities extracted from every pixel in the image [15]. Basically, in a color model such as *RGB*, there exist three color channels: red, green, and blue, in which each channel's size is 8 bits and therefore it ranges from 0 to 255 different intensities for the same color. In total, a pixel is 24 bits and thus can have a value of 0 representing complete black and 16,777,215 representing a complete white. The remaining values that fall in between represent the spectrum of colors. In *IR* applications, color intensities are usually encoded into a histogram [16] that spans over the whole pixels of an image. A color histogram represents the frequencies of every intensity color in the image. From a probabilistic viewpoint, a color histogram represents the probability mass function of the image intensities [16]. It is formally defined as:

$$h_{R,G,B} = N \cdot \Pr ob\{R = r, G = g, B = b\}$$

where r , g and b denote the three color channels in an *RGB* color model and *N* denotes the number of pixels in the image. Computationally, the color histogram is calculated simply by counting the number of pixels for each color.

Usually, when retrieving images based on their color histogram, similarity metrics are employed to measure the distance between the query image and every indexed image. They include but not limited to Histogram Intersection Distance [17], Euclidean Distance [18], Mahalanobis Distance [19], and Histogram Quadratic Distance [20]. The image having the smallest similarity value with respect to the query image is favored over other images in the database.

### *2.2.2 Texture Features*

Intuitively, a texture refers to the occurrence of a homogenous spatial pattern that exhibits structural arrangement of surfaces such as water, sky, fabric, bricks, heat, natural scenes, and specific modalities. Practically, textures are extracted using several techniques, the best known are based on statistical and spectral models [21]. Nevertheless, other types exist, which are heavily based on mathematical transformation models and they include Wavelet Transform [22], Fourier Transform [23], Discrete Cosine Transform (DCT) [24], and Gabor Transform [25]. The extraction of texture features is considered a computationally intensive task especially when a large number of images need to be processed. This is primarily due to image segmentation which consists of partitioning an image into regions with common spatial characteristics, each of which having a unique texture distinct from its adjacent textures.

#### *2.2.3 Shape Features*

Shape features depict a geometrical object in the image space having a specific boundary, size, orientation, content, and other properties [26]. A shape can be recognized by several descriptors including roundness, circularity, points of curvature, corners, planes, and turning angles. Two categories for shape features extraction exist: The first one is called Contour-based which captures spatial information on boundary points and lines such as shape of a tree, of a man, or of a vehicle. Typically, the Contour-based method can be implemented using a variety of techniques including Shape Signatures [27], Fourier Descriptors [28], and Curvature Scale Space [29]. The second category is called Region-based and it captures spatial information within a particular shape such as pupil inside an eye shape, or particles inside a rock shape. Moment Descriptors [30] and Grid Descriptors [31] are implementation techniques for Region-based shape features extraction.

## 3. The Proposed Hybrid IR Model

This paper proposes a new information retrieval model for indexing and retrieving web images embedded within *HTML* documents. The model is hybrid in that it handles images based on both their content (*CBIR*), and their textual metadata (keyword-based *IR*). In effect, graphical content is captured by extracting color features from the image and converting them into a color histogram; whereas, textual metadata are extracted from the text that encloses the image.

Furthermore, this paper presents a new term weighting scheme called *VTF-IDF* short for Variable Term Frequency-Inverse Document Frequency to weight the different terms of the captured metadata. It is a variable scheme because it weights terms based on their location in the original *HTML* document.

The proposed model comprises two major processes: The indexing process and the retrieval process. Figure 1 depicts both processes along with their flow of execution.

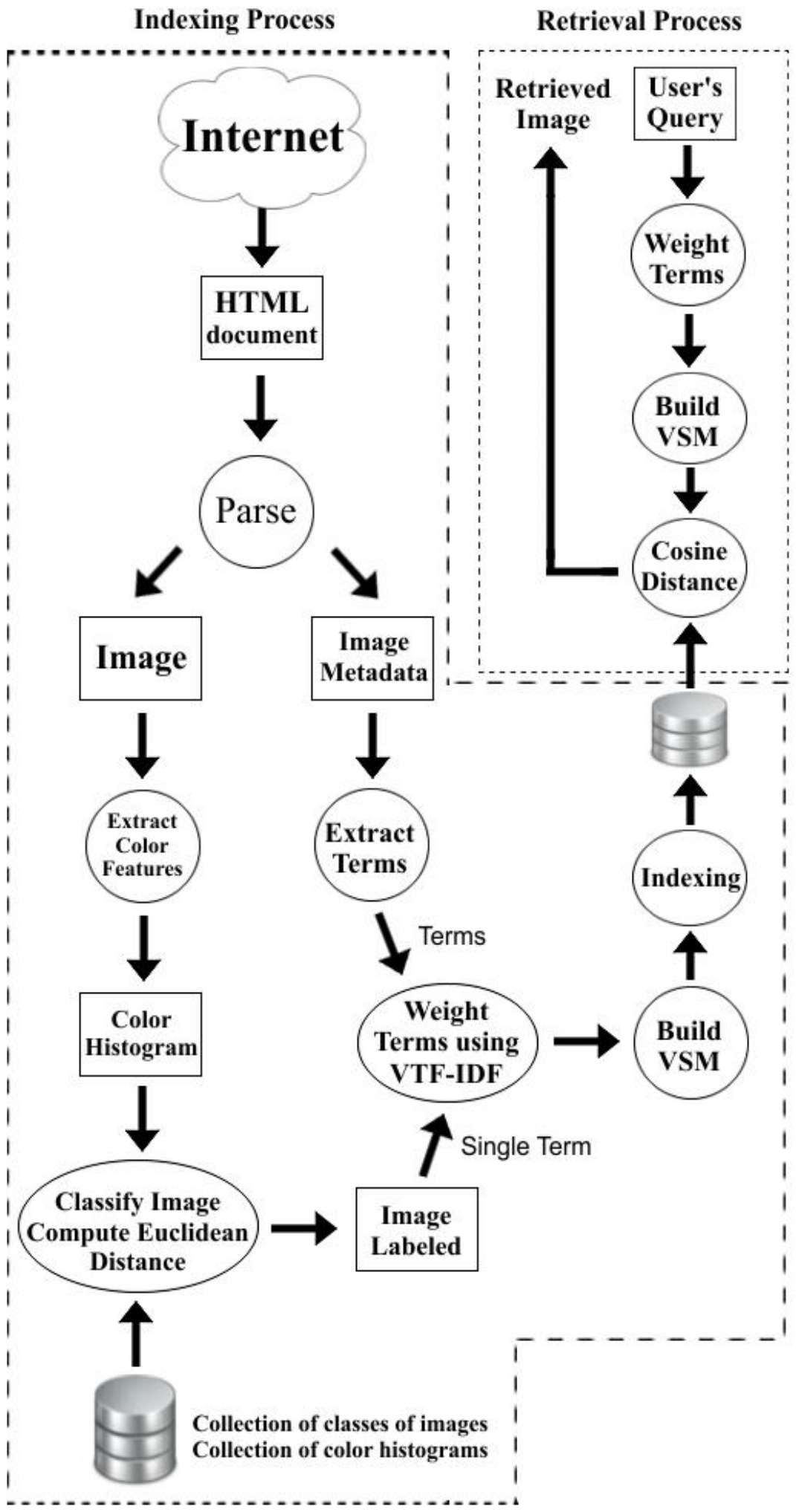

**Figure 1.** Execution flow for the proposed IR model

### 3.1 The Indexing Process

At the very beginning, a web image is fetched from the Internet, then its color features are extracted and converted into a 24-bit *RGB* 4-Bins color histogram in which bin 0 corresponds to intensities 0-63, bin 1 to 64-127, bin 2 to 128-191, and bin 3 to 192-255. Generally, the color histogram represents the number of pixels with same colors from each bin. The purpose of building the color histogram is to eventually classify the image. For this reason, the obtained color histogram is compared with other color histograms from the Corel image database [4] which houses a collection of preset images grouped into classes having similar color distribution. Then, the Euclidean distance is used to find how much the histogram of the image being indexed is close to other histograms in Corel database. As a result, the web image is tagged with the closest class in Corel database, denoted by *class-label*. The Corel classes are: Africa, Beach, Buildings, Buses, Dinosaurs, Elephants, Flowers, Food, Horses, and Mountains.

Mathematically, the Euclidean distance is defined as follows:

$$\sqrt{\sum_{i=1}^{n}(\mathbf{H1}_i - \mathbf{H2}_i)^2}$$

where *n* is the number of pixels in the histogram, *i* denotes a particular color intensity, *H1* denotes the color histogram of the image being indexed and *H2* denotes the color histogram of a particular image from the Corel collection.

Once the color histogram is built and the image *class-label* is recognized, the textual metadata are captured from the *HTML* text that surrounds the image. Actually, the extraction of metadata consists of pulling the terms (words) out of the text that wraps the image. Specifically, they include the terms of the *<p>* paragraph tag, the terms of the preceding *<h1>* and *<h2>* header tags, the terms of the image's *alt* attribute and *filename*, and finally the sole term of the *class-label*. Figure 2 shows the different metadata locations in the *HTML* document considered by the proposed model during the indexing phase.

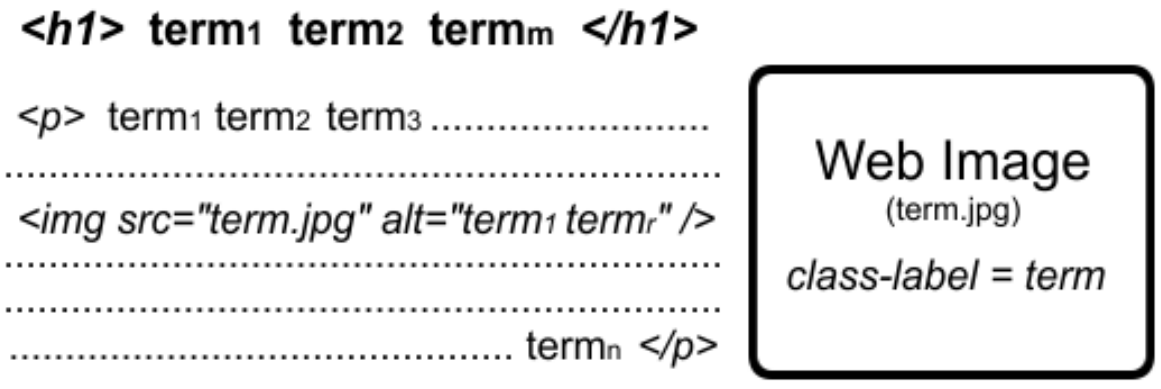

**Figure 2.** Metadata location

## 3.2 The VTF-IDF Term Weighting Scheme

In order to weight the extracted metadata terms, a new weighting scheme is proposed called *VTF-IDF* short for Variable Term Frequency-Inverse Document Frequency. Intrinsically, *VTF-IDF* is particularly designed to weight terms that surround web images embedded within *HTML* web documents. In other terms, *VTF-IDF* exploits the tag structure of the *HTML* language to deduce the magnitude of certain terms with regard to other terms, and consequently to assign them higher weights. For example, terms that appear in the *<p>* tag around the image should be weighted normally. However, terms that appear in the *alt* attribute and in the *<h1>* tag should be considerably weighted as they might be correlated to the semantics of the image. Likewise, terms that appear in the image's *filename* and its *class-label* are assigned the highest weight as they have the greatest semantic significance to describe the content of the image. Basically, *VTF-IDF* is derived from the classical *TF-IDF* weighting scheme, however, modified so that it assigns extra bonus weight for terms located within some special *HTML* tags, in addition for terms of the image's *alt* attribute, its *filename*, and its *class-label*. The *VTF-IDF* scheme can be formally defined as:

$$w_{ij} = (tf_{ij} * variable_weight) * idf_j$$

where $w_{ij}$ is the final weight for term *j*, $tf_{ij}$ is the frequency of term *j* enclosing image *i*, $idf_j$ is the Inverse Document Frequency of term *j*, and *variable_weight* is a number whose different values are outlined in Table 1.

**Table 1.** Different term locations and their corresponding weights

| Term Location | Variable Weight |
|---|---|
| <p> | 1 (normal weight) |
| <h1> | +10 |
| <h2> | +10 |
| <alt> | +10 |
| Image's filename | +20 |
| Image's class-label | +20 |

Ultimately, the set of all term weights are encoded into a feature vector based on the classical *VSM* (Vector Space Model) [5], and then indexed into a relational database for later retrieval. The *VSM* is algebraically defined as $img_j = \{w_{1,j}, w_{2,j}, w_{3,j}, w_{i,j}\}$ where *j* denotes a particular image, *i* denotes a particular distinct term enclosing $img_j$, and $w_{i,j}$ denotes the weight of the $i_{th}$ term enclosing image *j*. An example for a *VSM* representing a web image can be given as:

$$img = \{\ w(\text{hardware}),\ w(\text{PC}),\ w(\text{graphic}),\ w(\text{engine}),\ w(\text{pixel}),\ w(\text{acceleration}),\ w(\text{texel})\ \}$$

where $w(t_i)$ is a function that calculates the weight of term $t_i$ in image *img* using the *VTF-IDF* term weighting scheme. Supposing that the above terms have weights equal to 5, 3, 10, 10, 8, 14, and 1 respectively, then the results would yield to the following weighted *VSM*:

$$img = \{\ 5, 3, 10, 10, 8, 14, 1\ \}$$

### 3.3 The Retrieval Process

Initially, the retrieval process starts by a user submitting a text query *q* to the system. Subsequently, terms in the query are captured and weighted using the conventional *TF-IDF* weighting scheme. Then, the *VSM* is built for this query. Now, in order to compare between the query *q* and the collection of indexed images *img*, the cosine of the angles between vector *q* and every indexed vector *img* is to be calculated. When they are alike, they will receive a cosine of 1; when they are orthogonal (sharing no common terms), they will receive a cosine of 0 [32]. The image whose *VSM* yielded to the highest cosine value (close to 1) is returned to the user. More generally, the similarity between a vector $d_j$ and a vector *q* can be calculated using the following cosine equation:

$$sim(d_j, q) = \cos(\vec{d_j}, \vec{q}) = \frac{\vec{d_j} \bullet \vec{q}}{|\vec{d_j}||\vec{q}|} = \frac{\sum_{i=1}^{N} w_{i,q} \times w_{i,j}}{\sqrt{\sum_{i=1}^{N} w_{i,q}^2} \times \sqrt{\sum_{i=1}^{N} w_{i,j}^2}}$$

## 4. Experiments & Results

In the experiments, an extract is selected from a Wikipedia article [33] entitled “Blue Ridge Mountains”. It mainly encloses an RGB 24-bit image whose size is 384x256 pixels, its *filename* is “Rainy_Blue_Ridge.jpg”, and its *alt* attribute is “Blue Ridge”. Additionally, it encloses a paragraph *<p>* and a header *<h2>*. Figure 3 shows a screenshot of this extract.

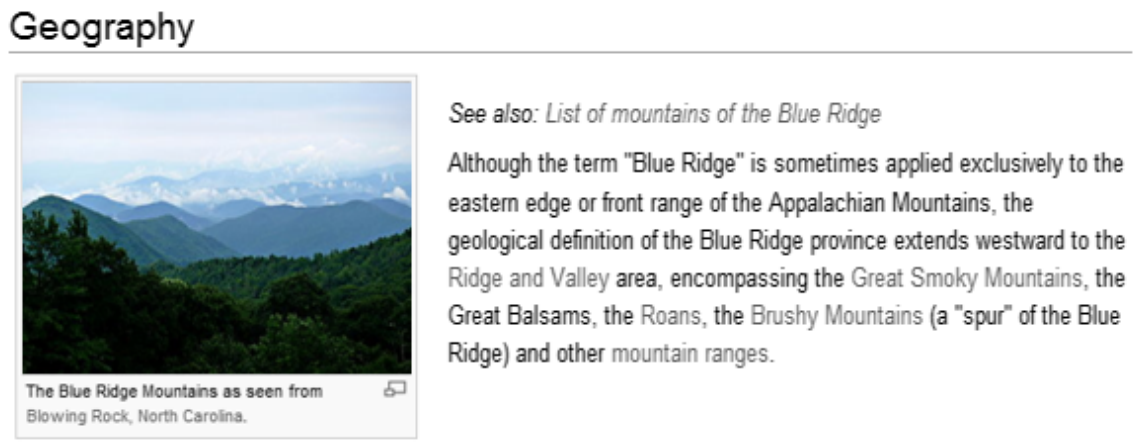

**Figure 3.** Extract from Wikipedia

### 4.1 The Indexing Process

The first step in the indexing process is to connect to the selected Wikipedia article webpage and download the particular web image. The second step is to extract the image’s color features and built its color histogram. Table 2 shows the color histogram of the downloaded web image organized into 4 bins, along with the number of pixels in each particular bin.

**Table 2.** Color Histogram

| **Red** | | | |
|---|---|---|---|
| Bin 0 | Bin 1 | Bin 2 | Bin 3 |
| 46508 | 11207 | 6928 | 33661 |
| **Green** | | | |
| Bin 0 | Bin 1 | Bin 2 | Bin 3 |
| 39967 | 13387 | 6880 | 38070 |
| **Blue** | | | |
| Bin 0 | Bin 1 | Bin 2 | Bin 3 |
| 41811 | 4328 | 10038 | 42127 |

After building the color histogram, the Euclidean distance to other histograms in the Corel database is computed. The image being indexed is then classified under the class of the closest histogram. Table 3 shows the results of this computation. The Corel “Mountains” class received the closest average Euclidean distance and thus the *class-label* of the Wikipedia image “Rainy_Blue_Ridge.jpg” is set to “Mountains”.

**Table 3.** Results of calculating the Euclidean distance

| **Corel Classes 10 classes with 100 images each** | **Average Euclidean Distance** $Ed(H1,H2) = \sqrt{\sum_{k=1}^{n} (H1_k - H2_k)^2}$ Avg $Ed(H1,H2) = (\sum_{k=0}^{100} Ed) / 100$ |
|---|---|
| Africa | 60074.60 |
| Beach | 54441.60 |
| Buildings | 65280.47 |
| Buses | 58400.45 |
| Dinosaurs | 68935.05 |
| Elephants | 58804.01 |
| Flowers | 70700.41 |
| Food | 68480.77 |
| Horses | 71664.69 |
| **Mountains** | **47162.84** |

As the *class-label* was defined, the next step is to extract the textual metadata from the surrounding text of the image. The first two columns in Table 4 list the extracted metadata terms along with their original location in the *HTML* document. It is worth noting that the metadata was post-processed in order to remove all occurrences of stop words.

The following step is to weight the extracted terms using the proposed *VTF-IDF* term weighting scheme. The complete computation together with the resulted weights are all given in Table 4. It is worth noting that the Inverse Document Frequency (*IDF*) is considered equal to 1 as for illustration purposes only one single image is being indexed and not a collection of images.

**Table 4.** Metadata terms and their weights

| Term | Location | Freq | VTF |
|---|---|---|---|
| Mountain | Class-Label | 1 | 1 * *variable_weight(class-label) = 20* |
| | <p> | 5 | 5 * *variable_weight(<p>) = 5* |
| | | | ***Total = 20+5=25*** |
| Rainy | File-name | 1 | 1 * *variable_weight(filename) =* ***20*** |
| Blue | File-name | 1 | 1 * *variable_weight(filename) = 20* |
| | Alt | 1 | 1 * *variable_weight(alt) = 10* |
| | <p> | 4 | 4 * *variable_weight(<p>) = 4* |
| | | | ***Total = 20+10+4=34*** |
| Ridge | File-name | 1 | 1 * *variable_weight(filename) = 20* |
| | Alt | 1 | 1 * *variable_weight(alt) = 10* |
| | <p> | 5 | 5 * *variable_weight(<p>) = 5* |
| | | | ***Total = 20+10+5=35*** |
| Geography | <h2> | 1 | 1 * *variable_weight(<h2>) =* ***10*** |
| Range | <p> | 2 | 2 * *variable_weight(<p>) =* ***2*** |
| Great | <p> | 2 | 2 * *variable_weight(<p>) =* ***2*** |
| See, Also, List, Although, Term, Sometimes, Applied, Exclusively, Eastern, Edge, Front, Appalachian ,Geological, Definition, Province, Extends, Westward, Valley, Area, Encompassing, Smoky, Balsams, Roans, Brushy, Spur, Other | <p> | 1 for each term | 1 * *variable_weight(<p>) =* ***1*** |

The final step is to build the Vector Space Model (*VSM)* of the image which encodes all computed terms weights. The actual *VSM* is given below:

*img* = { *w*(Mountain) , *w*(Rainy) , *w*(Blue) , *w*(Ridge) , *w*(Geography) , *w*(Range) , *w*(Great) , *w*(See) , *w*(Also) , *w*(List) , *w*(Although) , *w*(Term) , *w*(Sometimes) , *w*(Applied) , *w*(Exclusively) , *w*(Eastern) , *w*(Edge) , *w*(Front) , *w*(Appalachian) , *w*(Geological) , *w*(Definition) , *w*(Province) , *w*(Extends) , *w*(Westward) , *w*(Valley) , *w*(Area) , *w*(Encompassing) , *w*(Smoky) , *w*(Balsams) , *w*(Roans) , *w*(Brushy) , *w*(Spur) , *w*(Other) }

*img* = { *25 , 20 , 34 , 35 , 10 , 2 , 2 , 1 , 1 , 1 , 1 , 1 , 1 , 1 , 1 , 1 , 1 , 1 , 1 , 1 , 1 , 1 , 1 , 1 , 1 , 1 , 1 , 1 , 1 , 1 , 1 , 1 , 1*}

## 4.2 The Retrieval Process

Assuming that a user has submitted a query *q*="mountain blue ridge" during the normal operation of the system. The first step is to model this query into a *VSM*, then using the traditional *TF-IDF* weighting scheme to weight its enclosing terms namely "mountain", "blue", and "ridge" respectively. The actual weighted *VSM* for query *q* is given below:

*q* = ( *w*(*mountain*), *w*(*blue*), *w*(*ridge*) )
*q* = ( *1, 1, 1* )

In order to find if query *q* has a match, the similarity between the two vectors, namely the query *q* and every indexed image *img* is computed using the cosine metric. The image *img* having the highest cosine value is considered as relevant. It is worth mentioning that the normalization of the two vectors is integrated directly into the similarity equation. The complete math for calculating the cosine metric between the query *q* and the already indexed *VSM* of the initial "Rainy_Blue_Ridge.jpg" image (represented in the equation as *d*) is sketched below:

$$sim(d_j, q) = \cos(\vec{d_j}, \vec{q}) = \frac{\vec{d_j} \bullet \vec{q}}{|\vec{d_j}||\vec{q}|} = \frac{\sum_{i=1}^{N} w_{i,j} \times w_{i,q}}{\sqrt{\sum_{i=1}^{N} w_{i,j}^2} \times \sqrt{\sum_{i=1}^{N} w_{i,q}^2}}$$

$$|\vec{d_j}| = \sqrt{\sum_{i=1}^{N} w_{i,j}^2} = \sqrt{\begin{matrix}25^2+20^2+34^2+35^2+10^2+2^2+2^2+1^2+1^2+1^2+1^2+1^2+1^2+1^2+1^2+ \\ 1^2+1^2+1^2+1^2+1^2+1^2+1^2+1^2+1^2+1^2+1^2+1^2+1^2+1^2+1^2+1^2+1^2+1^2\end{matrix}}$$

$$= \sqrt{3{,}540} = 59.49$$

$$|\vec{q}| = \sqrt{\sum_{i=1}^{N} w_{i,q}^2} = \sqrt{1^2+1^2+1^2}$$

$$= \sqrt{3} = 1.73$$

$$\vec{d_j} \bullet \vec{q} = 25{*}1 + 34{*}1 + 35{*}1 = 94$$

$$\frac{\vec{d_j} \bullet \vec{q}}{|\vec{d_j}||\vec{q}|} = 94 / (59.49 * 1.73) = 0.91$$

The above calculations yielded to a cosine value equal to 0.91. Fundamentally, when two vectors are alike, they will receive a cosine of 1; when they are orthogonal (sharing no common terms), they will receive a cosine of 0 [32]. Therefore, the above results indicate a 91% relevancy between the query *q* and the image *img* (represented in the equation as *d*) which is originally the web image "Rainy_Blue_Ridge.jpg" from the Wikipedia article.

## 5. Evaluation & Comparison

A head-to-head comparison was performed between the proposed hybrid *IR* model and between other *CBIR* engines available on the market. They are respectively Bing [34], Google [35], Gazopa [36], AltaVista Photofinder [37], Incogna [38], and Tiltomo [39]. The Precision metric was used to measure how much the returned images were relevant to the query. Primarily, it is calculated using the following equation [40].

$$Precision = \frac{\text{\# of relevant documents returned}}{\text{\# of documents returned}}$$

Figure 4 outlines the obtained results using a graphical histogram. In this figure, it is evident that the proposed *IR* model outperformed the other models as it scored the highest Precision rate. Obviously, around 70% of the returned images were relevant using the proposed model; while the best of the rest namely Gazopa nearly scored 51%. As a result, 19% more relevant images were retrieved by the proposed *IR* model.

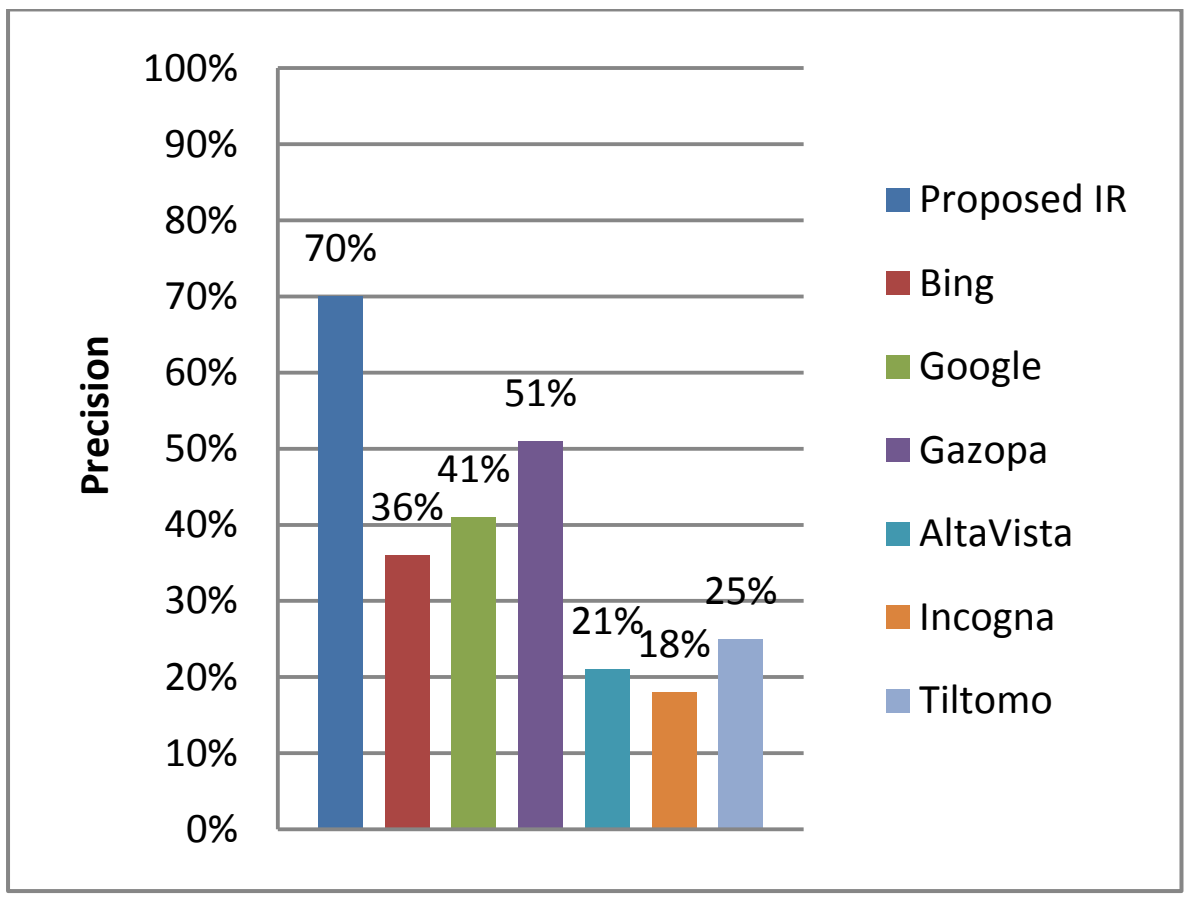

**Figure 4.** Precision metric for different models

## 6. Conclusions and Future Work

This paper presented an original image information retrieval model to index and retrieve web images embedded within *HTML* documents. The model uses *CBIR* techniques to retrieve images based on their content and keyword-based *IR* techniques to retrieve images based on their enclosing textual metadata. More to the point, terms of the textual metadata were weighted using a novel term weighting scheme called *VTF-IDF* which assigns variable weights for terms depending on the *HTML* tags they appear in. Consequently, this combination of using graphical content alongside with textual metadata, in addition to a weighting scheme that evaluates terms according to their semantic significance with respect to the *HTML* tag structure, has led to high image retrieval Precision rate that outperformed other traditional and existing image *IR* models.

As future work, the proposed image *IR* model can be improved so much so it supports image retrieval based on their texture features allowing searching for visual objects and specific content within images such as finding all images in a larger set of images which have a tree or a person. Furthermore, the proposed model is to be extended to support the retrieval of video and audio clips so as to provide the capability of searching for multiple types of web multimedia files.

## Acknowledgments

This research was funded by the Lebanese Association for Computational Sciences (LACSC), Beirut, Lebanon under the "Web Information Retrieval Research Project – WIRRP2011".